\documentclass[aip,rsi,amsmath,amssymb,reprint]{revtex4-1}

\usepackage[hmargin=1.9cm,vmargin=2cm]{geometry}
\usepackage{graphicx}
\usepackage{dcolumn}
\usepackage{bm}

\usepackage{caption}
\newcommand{\ket}[1]{| {#1} \rangle}

\begin{document}


\title[FPGA-Controlled Versatile Microwave Source for Cold Atom Experiments]{FPGA-Controlled Versatile Microwave Source for Cold Atom Experiments}

\author{Isaiah Morgenstern}
\author{Shan Zhong}
\author{Qimin Zhang}
\author{Logan Baker}
\author{Jeremy Norris}
\author{Bao Tran}
\affiliation{ 
The University of Oklahoma, Homer L. Dodge Department of Physics and Astronomy, 440 W. Brooks St, Norman, Oklahoma 73019, USA
}%
\author{Arne Schwettmann}
\email{schwettmann@ou.edu}
\affiliation{ 
The University of Oklahoma, Homer L. Dodge Department of Physics and Astronomy, 440 W. Brooks St, Norman, Oklahoma 73019, USA
}%


\begin{abstract}
We present a microwave source that is controlled by a commercially available field programmable gate array (FPGA). Using an FPGA allows for precise control of the time dependent microwave-dressing applied to a sample of trapped cold atoms. We test our microwave source by exciting Rabi oscillations in a Na spinor Bose-Einstein Condensate. We include, as supplements, the complete source code, parts lists, pin connection diagrams, and schematics to make it easy for any group to build and use this device.
\end{abstract}

\maketitle

\section{Introduction}

Microwave-dressing can be used in atomic physics experiments to apply shifts to hyperfine energy levels with many applications. For example, using microwave fields, it is possible to control spin-exchange collisions in an ultracold sodium gas by changing the effective quadratic Zeeman shift through the AC Stark effect.\cite{Liu2014} This technique was recently used to implement a phase-sensitive amplifier for SU(1,1) interferometry.\cite{Lett2018}

In this paper, we describe the implementation of a microwave source controlled by a field-programmable gate array (FPGA), including, in the supplements, the source code, parts lists, pin connection diagrams, and detailed instructions. Using an FPGA has many advantages, one being easy re-programming so that the function of the device can be adapted to changing experimental needs, including servo controllers and other functions as needed.\cite{Shaffer2011,Madison2018}

FPGA devices are becoming more common in experimental physics because of their versatility and wide range of functions.\cite{Shaffer2011, Madison2018, Salapaka2018, Katori2015, Amini2013} An FPGA offers a compact and controllable way to manage experimental equipment without the need to devote extra devices, because an FPGA can run several tasks in parallel without any slow-down. FPGAs are also rated for various lab conditions; some FPGAs have been shown to work in 4 K with minimal errors.\cite{Reilly2016} This allows the FPGA to be in the same enclosure as the device or detector, removing the physical barrier and limiting electrical losses from separation.

In this article, we present a versatile FPGA-based microwave source. This source is used to control spin-mixing collisions in our Na spinor BEC with the goal of implementing a spin-mixing interferometer and other devices in matter-wave quantum optics.\cite{Schwettmann2019} We use an FPGA together with a direct digital synthesizer (DDS) and frequency mixing at GHz frequencies. This combination allows for the production of microwave and radio frequency pulses with controllable parameters and timing sequences. The frequency, amplitude, and phase of the waves are set and programmed into the DDS by the FPGA. Parameters can be changed quickly with 5 $\mu s$ timing resolution by storing pre-programmed timing sequences on the FPGA board memory. The FPGA steps through the timing sequence one-step at a time on receipt of an external TTL clock pulse. The timing sequence is written to the FPGA board memory via a desktop or laptop computer connected via USB, running LabVIEW. The FPGA receives timing sequences while it is running. This allows for the timing sequences to be designed in LabVIEW and executed by the FPGA without having to reprogram the FPGA every time the sequence changes. We designed our microwave source to create microwaves at 1.77 GHz, close to the $F=1$ to $F=2$ ground state hyperfine transitions in sodium. This frequency is needed in our experiments to control spin-mixing collisions in a sodium spinor Bose-Einstein Condensate (BEC). To test the functionality of our microwave source, we use it to excite and measure Rabi oscillations between the $\ket{F,m_F}=\ket{1,0}$ and the $\ket{2,0}$, $\ket{2,1}$ and $\ket{2,-1}$ states in our BEC. In our tests, we apply a small dc magnetic field to lift the degeneracy of these states via the Zeeman effect.

\section{Device}

In the following sections, we describe our device in detail, including the hardware, software and operation. Further details, such as pin layouts, source codes, FPGA programming files, and installation instructions can be found in the supplements.\cite{Supplements}

\subsection{Commercial Boards}

\begin{figure}[ht!]
\includegraphics[width=2.5in]{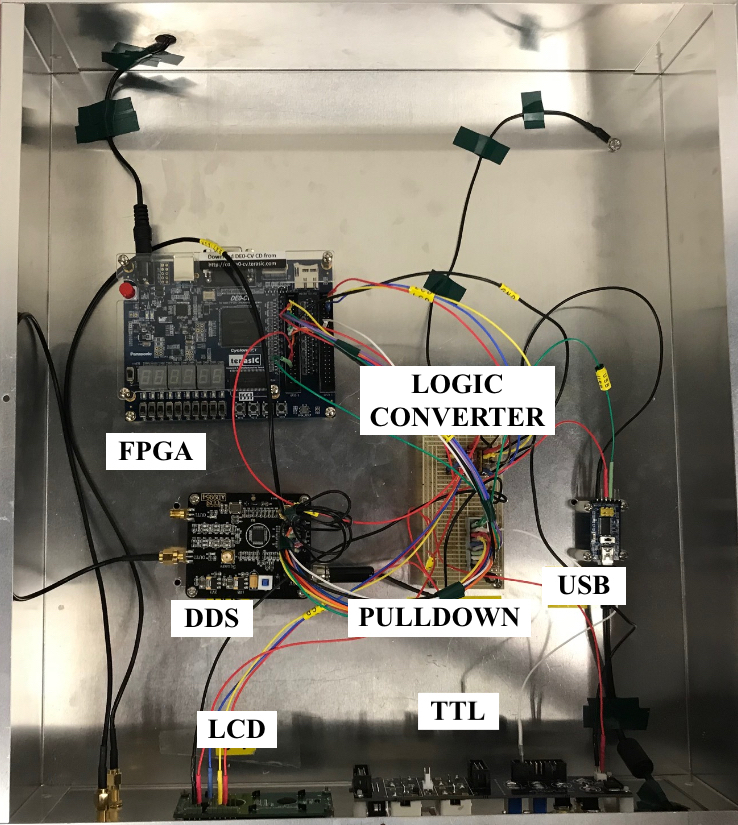}
\caption{(Color online) Top-down photo looking at all the boards in the enclosure.}
\label{fig_1}
\end{figure}

Our FPGA board (Terasic DE0-CV) houses an Altera Cyclone V FPGA chip along with two input/output headers. No modifications to the board are required to implement our design, but a few pulldown resistors need to be connected to some of the inputs. The DDS and input TTL signal are connected via 100 $\Omega$ pull-down resistors to ground. The USB input is connected via a 1~M$\Omega$ pulldown resistor to ground. The FPGA board also houses a flash memory chip (Altera EPCS64) that stores the FPGA program. If the FPGA ever loses power, it does not have to be reprogrammed. The functionality will be restored when power is returned. Using a commercial board saves the hassle of reflow soldering and device testing associated with using a custom board.

Along with the FPGA board, our microwave source consists of three other commercial boards: a DDS board (Diybigworld AD9954 DDS DWORLDS-2DS63), a USB input board (DWS Q90 New Basic FTDI FT232) for receiving data, and an LCD screen (Newhaven Display NHD-0216K3Z-FL-GBW-V3) to aid with user interaction via the front panel. These boards are enclosed in a grounded aluminum box (Bud Industries AC-1428). The FPGA, DDS, and USB boards are secured within the center of the box, see Fig.~\ref{fig_1}.

\begin{figure*}[ht!]
\includegraphics[width=\textwidth]{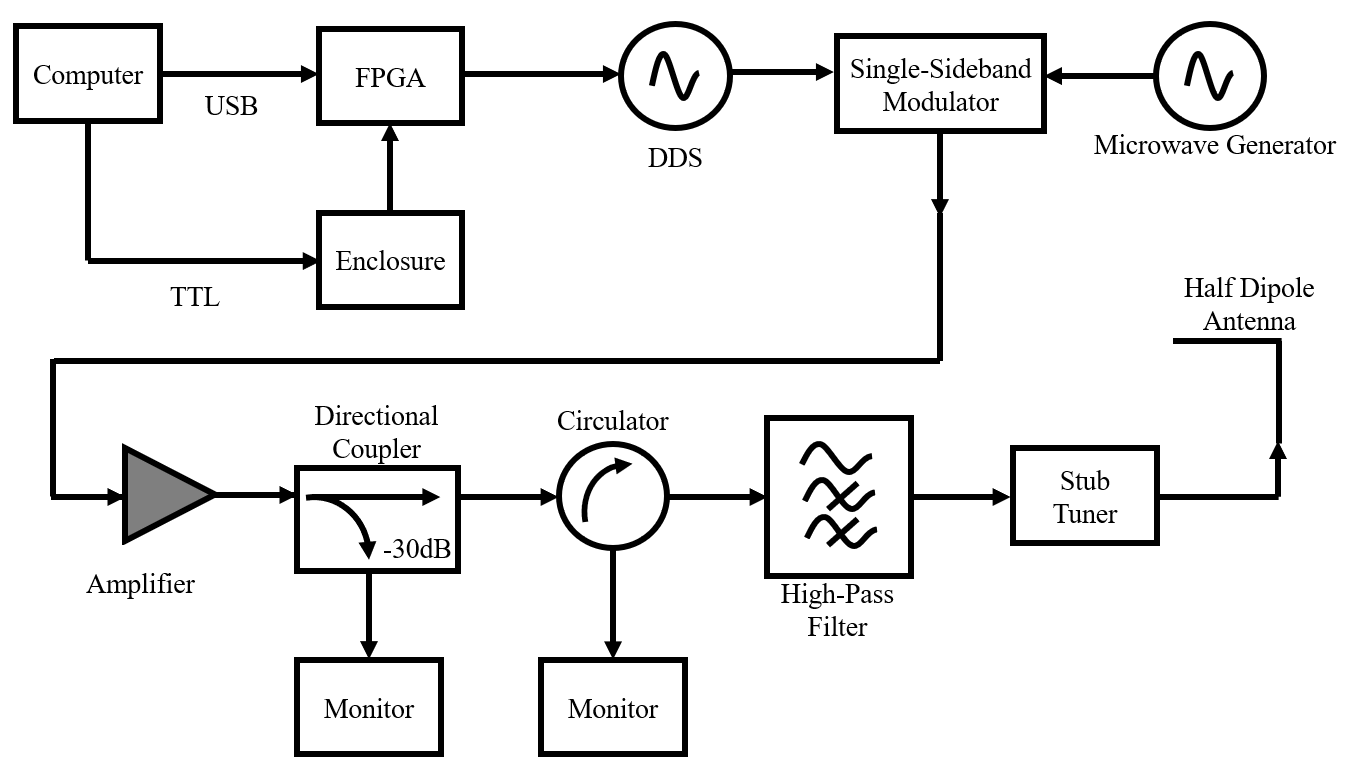}
\caption{Schematic of microwave signal generation hardware. The parameters for the low frequency signal are sent from the computer to the FPGA via USB. The FPGA then communicates with the DDS which produces a low-frequency (MHz) signal. The low-frequency signal is mixed with a high-frequency signal of 1.73 GHz in a single-sideband modulator. The modulator produces the desired sum-frequency signal close to the hyperfine splitting of the sodium ground state. After amplification, filtering, and impedance matching, the signal is then emitted from a half dipole antenna to irradiate a sample of trapped cold atoms.}
\label{fig_2}
\end{figure*}

\subsection{Device Operation}

Switches and buttons are built into the FPGA board and accept user inputs. A switch is used to change between manual and remote mode. Manual mode is used to test that the DDS and FPGA are working properly. In manual mode, a rudimentary menu is displayed on the LCD screen, controlled via up, down, and ok buttons on the FPGA board. To operate manual mode three of the buttons are used: switch parameter, value up, and value down. The switch parameter button cycles between DDS parameters. The value up and value down buttons adjust the selected parameter accordingly. As the user switches between the parameters, the LCD screen will display the selected parameter and its value. The menu allows setting amplitude and frequency. Any change of these parameters causes the output to update immediately. In remote mode, the FPGA listens for timing sequence data via the USB connection. Once the timing sequence data is received, it is stored in volatile memory on the FPGA board. When the FPGA receives a rising TTL trigger on one of the inputs, it will step to the next set of parameter values (amplitude, frequency and phase) in the timing sequence table and updates the DDS output accordingly. The FPGA receives TTL signals via a BNC connector on the enclosure, connected to one of the input pins of the FPGA board (see Supplemental Materials for pin layouts). In our setup, the TTL signals are generated by a digital output PCI card (Spincore Pulseblaster PB24-100-4k) inside the same computer that communicates with the FPGA via USB, but they can also be generated by any other function generator or data acquisition board. When in remote mode, the LCD screen just displays the words "REMOTE MODE" to indicate that manual user input is disabled.

The LCD board uses 5V logic while the FPGA uses 3.3V logic. To work around this issue, we employed a logic level converter (SparkFun Logic Level Converter - Bi-Directional BOB-12009) when connecting the LCD panel to the FPGA board. The low voltage side of the logic level converter has 100 $\Omega$ pull-up resistors connected to 3.3V to ensure that the logic signal is high enough and no data bits are ever lost.

The DDS board is controlled by the FPGA board. All inputs into the DDS headers were connected to 100 $\Omega$ pull-down resistors to prevent floating voltages. There are several memory addresses in the FPGA that store frequency, amplitude and phase values for use as steps in the timing sequence. These data are stored as words in a table. When the FPGA receives a TTL signal it steps to the next memory block in the table and sends the corresponding parameters to the DDS chip, so that it can update the output signal. We use an external 400 MHz clock, generated by a stable function generator (Hewlett Packard HP8657B), connected to the external clock input on the DDS board. This enables the DDS to produce a change of output amplitude, frequency, or phase within four microseconds of receiving new parameters. The external clock also allows us to bypass the internal clock multiplier on the DDS chip to remove unwanted sidebands on the output signal. The output of the DDS is passed through a DC block (Mini-Circuits BLK-89-S+) to prevent any DC bias on the output signal.

A desktop computer is operating on Windows 10 and is running LabVIEW. LabVIEW communicates directly with the FPGA board via a USB connection. Through this connection, LabVIEW transmits timing sequences, which are stored on the FPGA board. The timing sequences are transmitted as strings that always start with the letter L. Thus, when the FPGA receives a sequence starting with L it decodes the following characters into amplitude, frequency, and phase parameter values that can be sent to the DDS.

Although the DE0-CV comes equipped with a USB connector, its use is limited only to reprogram the board. It cannot be used to communicate with the FPGA while it is running. Therefore, to allow receiving timing sequence data from LabVIEW, we added an external USB communication module. The module has several pins, of which only three are used for our purposes. The 5 V and ground pin provide power to the USB FTDI chip. The TxD (transmitted data) is used to transmit the data received from the computer via the USB protocol to the FPGA board. The connection between the USB board and the FPGA board is serial (COM). The CTS (clear to send), DTR (data terminal ready), RTS (request to send), and RxD (received data) pins are not utilized here.

\begin{figure*}[t]
\includegraphics[width=5in]{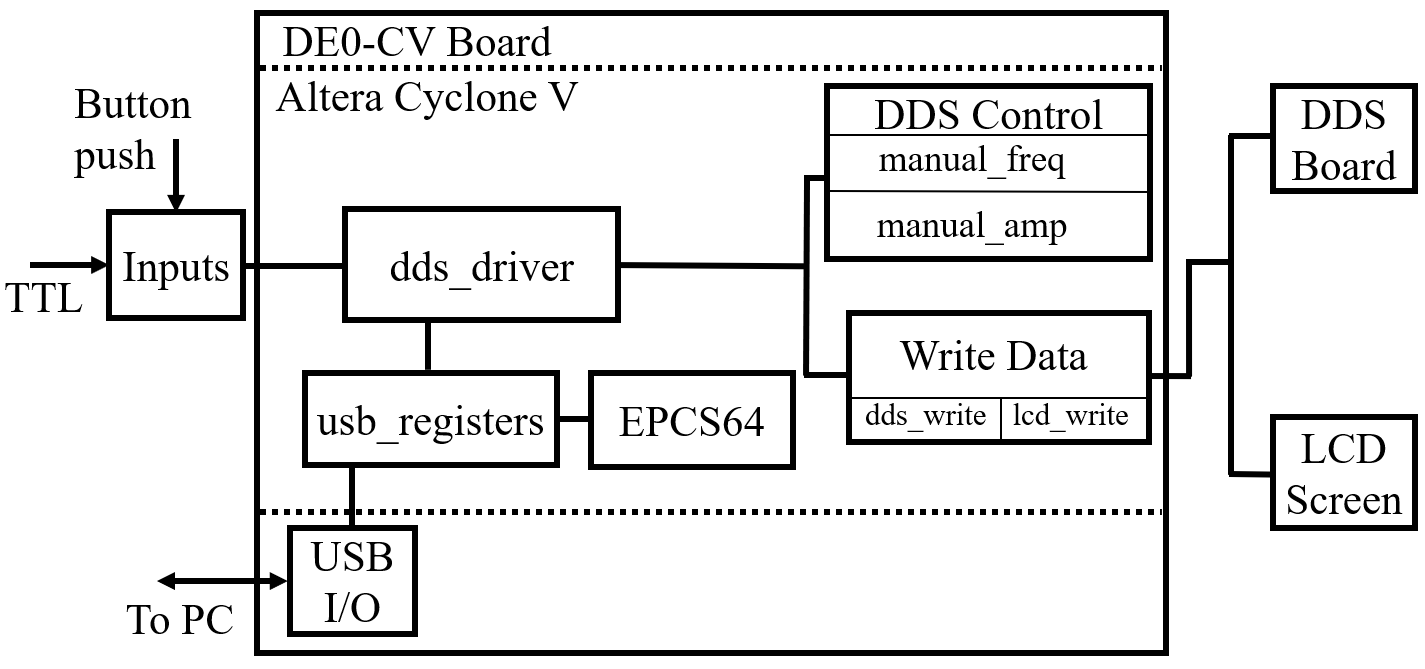}
\caption{Schematic of FGPA program structure. The dds\_driver is the main module that controls the device. The insets in DDS Control and Write Data denote separate Verilog modules. Usb\_registers is a sub-module of dds\_driver, used to implement communication with the computer.}
\label{fig_3}
\end{figure*}

\subsection{Signal Generation}

Fig. \ref{fig_2} depicts a simplified diagram of the microwave signal generation path. The FPGA board, as it steps through the timing sequence, sends commands to the DDS board to create a sine wave of the specified frequency and amplitude that is around 30 MHz in our experiments. The output from the DDS board is mixed with a constant high frequency signal of 1.7416 GHz from a signal generator (Hewlett Packard HP8657B). The mixing is done with a single-sideband modulator (Polyphase Microwave SSB0622A) which passes only the sum of the two input frequencies, resulting in the desired microwave signal which is only a few tens of kHz detuned from the ground state F=1, m=0 to F=2, m=0 clock transition in sodium at 1.7716 GHz. The detuning from the clock transition is then conveniently controlled by changing the DDS parameters only, keeping the HP signal generator at a fixed frequency.

The microwave is sent through an amplifier (HD Communications HD28747) with a maximum power of 20 W to increase the coupling to the atoms and increase on-resonance Rabi frequencies into the kHz range. In our setup,  this allows fast manipulation of spin populations and phases on the timescale of $\mu s$. After the amplifier, the signal is sent through a directional coupler (Pasternack PE2201-30) with the -30 dB output used to monitor the signal on a spectrum analyzer.

Following the directional coupler is a circulator (DITOM D3C0120), used to monitor back reflections from the antenna and prevent such reflections from coupling back into the amplifier. This is important because any high power back reflections could destroy the amplifier. A stub tuner (Maury Microwave 1819B) is used to match impedance to our home-built half dipole antenna. To match the impedance the stub tuner is adjusted to minimize the observed back reflections on the third port of the circulator. This is done at low power, to prevent the antenna, which is inside our vacuum chamber from warming up and outgassing. If the antenna is run at high power for $\sim 100$ ms it will begin to outgas and affect the ultra-high vacuum (UHV). To prevent this, an interlock was installed that shuts down the amplifier if the chamber pressure rises. Inbetween the circulator and the stub tuner is a high-pass filter (Mini-Circuits SHP-1000+) 1000-3000 MHz, to filter out any potential unwanted dc levels or harmonics. The high-pass filter is directly connected to the antenna which is discussed in the experimental setup section later.

\begin{figure*}[t]
\includegraphics[width=\textwidth]{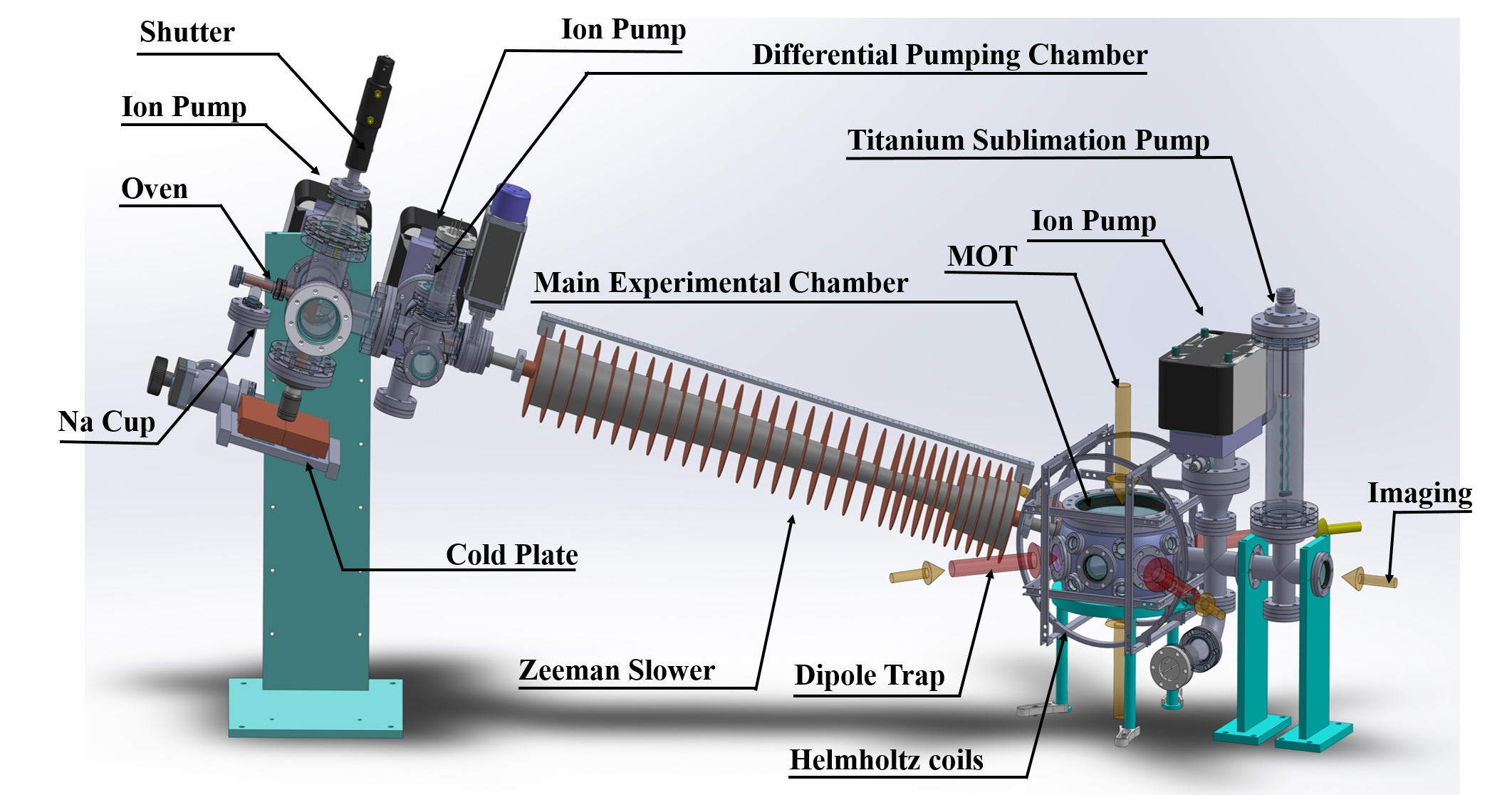}
\caption{(Color online) Schematic of main experimental setup. In the oven and differential pumping chamber, a nozzle, cold plate, and several collimation tubes create a collimated atomic beam of Na atoms. A pneumatic shutter is used to control the atomic beam. The Zeeman slower pre-slows the atomic beam to velocities of $\sim$20 m/s, before it enters the main vacuum chamber. The atomic beam is used to load a magneto-optical trap in the chamber center. A far-off resonance trap in crossed-beam geometry is then loaded from the magneto-optical trap. In the far-off resonance trap, evaporative cooling is used to cool the gas below the transition temperature of $\sim$100 nK and create a Bose-Einstein condensate. Detection of spin populations is done via Stern-Gerlach time-of-flight absorption imaging along the depicted imaging beam direction.}
\label{fig_4}
\end{figure*}

\section{Verilog Code}

Fig.~\ref{fig_3} shows a simplified block diagram of the Verilog program for controlling the DDS. The DE0-CV buttons provide signals between 0 V and 3.3 V to change the parameters of the DDS output in manual mode or step to the next part of the timing sequence in remote mode. The button voltages are passed directly to the Cyclone V FPGA chip. They are handled in the dds\_driver module, which is the main module of the code. The dds\_driver calls all other modules and passes data between them for the generation of the DDS control commands. The dds\_driver also receives timing sequence data from the USB connection. These data are written to registers in the EPCS64 memory chip on the DE0-CV board.

The dds\_driver controls two different groups of sub-modules. A DDS control group and a data writing group. In the data writing group, there are again two sub-modules: lcd\_write and dds\_write. The lcd\_write writes data to the LCD screen. For example, in remote mode it simply causes the LCD screen to display "REMOTE MODE." In manual mode this module will pass the values of the selected parameter to the LCD screen to allow user navigation of the menu. The dds\_write writes data to the DDS chip to control the frequency, amplitude and phase of the output signal. When a parameter is changed by a button push it signals the dds\_driver module to call both the lcd\_write and dds\_write to update both the DDS as well as the LCD screen. In remote mode, when the dds\_driver receives a TTL pulse, it steps to the next set of parameters in the timing sequence, retrieving data from the appropriate register in the EPCS64. Then, it calls dds\_write to pass this data to the DDS chip.

The DDS control block contains all the submodules called by dds\_driver when the device is in manual mode. Most of these submodules control the parameters of the DDS output. The manual\_freq and manual\_amp modules control the frequency and the amplitude of the DDS output respectively. These modules are called when a button on the DE0-CV board is pressed. The modules change the parameter and store the new value on the FPGA board while sending the change to the DDS. The modules also convert the binary representation of the value to a binary-coded decimal value which is then sent to the LCD screen for display.

\section{Experimental Setup}

In the following sections, we describe the experimental setup where we implemented our microwave source.

\subsection{BEC}
A schematic of our experimental setup is depicted in Fig. \ref{fig_4}. In our experiment, we produce an all-optical Na spinor BEC using laser-cooling and trapping and evaporative cooling. We then examine the time evolution of the spin populations of the $\ket{1, m_F}$ ground state sublevels as the atoms undergo spin-changing collisions. We heat Na atoms in an oven that is connected to the UHV system via a thin and long heated copper nozzle. After the beam exits the nozzle, a thick copper plate, cooled to $\sim-10\;^{\circ}\textrm{C}$, with a small hole in the center, helps to remove sodium atoms exiting at larger angles, and keeps the pressure in the oven chamber below $10^{-7}\;\textrm{torr}$. The nozzle and cold plate produce a collimated atomic beam. Behind the cold plate is a pneumatic shutter which is used to block the atomic beam when it is not needed. The atomic beam passes through a differential pumping chamber, and a Zeeman slower where atoms are pre-slowed to $\sim 20$ m/s. After the Zeeman slower, the beam enters the main experimental chamber, which is a extended spherical octagon with eight 2.75 CF ports, two 8 CF ports and sixteen 1.33 CF ports (Kimball Physics MCF800-ExtOct-G2C8A16). In the experimental chamber, a magneto optical trap (MOT) is loaded from the beam. After several seconds, when the MOT has fully loaded, a $30\;\textrm{ms}$ polarization-gradient cooling and optical molasses phase reduces the temperature to $\sim80\;\mu \textrm{K}$. We then directly load atoms into a crossed-beam far-off resonance trap (FORT) formed by two focused 1064 nm laser beams that overlap in the highest-density region in the center of the MOT. We achieve a Na spinor BEC via a 5 second long forced evaporative cooling sequence by lowering the FORT single-beam power from $\sim 20\;\textrm{W}$ to less than 1 W. Detection of spin populations is done via Stern-Gerlach time-of-flight absorption imaging using a CCD camera.

\subsection{Microwave Source}

Our microwave source setup is separate from the main experimental setup seen in Fig.~\ref{fig_4}. The microwaves are radiated inside the main experimental chamber by a home-built half-dipole antenna. The microwave source connects to the antenna through an electrical SMA feedthrough located on one of the 1.33 CF ports of our experimental chamber.

\begin{figure}[h!]
\includegraphics[width=180pt]{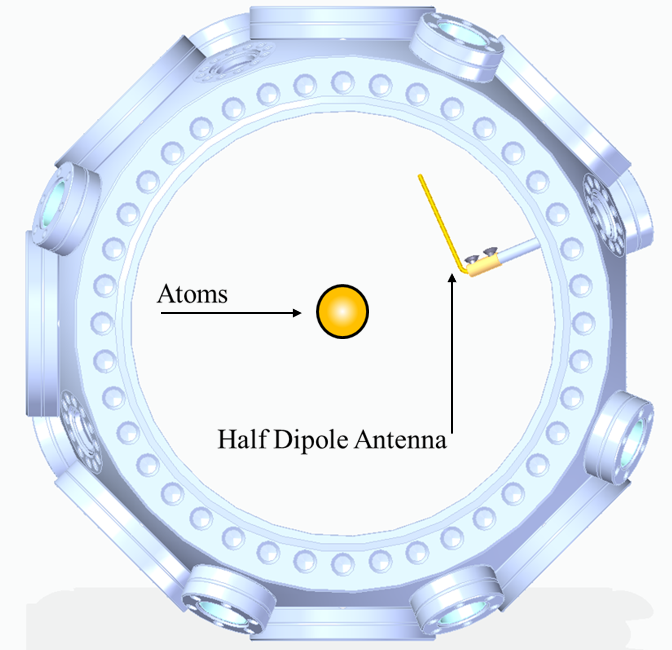}
\caption{(Color online) Sketch of the half-dipole (quarter-wave) antenna inside our stainless steel vacuum chamber. The antenna is oriented parallel to the atoms, because the microwaves radiate in all azimuthal directions. The antenna is approximately 1.2 inches away from the center of the cloud of trapped atoms. It is mounted on a UHV compatible coax cable connected to a 1.33 CF electrical feedthrough. Other viewports can be seen on the edges of the chamber, used for the MOT beams, FORT beams, imaging beams, Zeeman slower, and pump connections.}
\label{fig_5}
\end{figure}

\subsection{Antenna}

Our half-dipole antenna is a quarter wave antenna. The length of the antenna was chosen to be 4.2 cm to match our frequency of 1.771 GHz. The antenna is inside the UHV chamber. It is mounted to the chamber on an electrical feedthrough (Kurt J. Lesker IFTCG012012). The feedthrough connects to the antenna via UHV compatible kapton coax wire (Kurt J. Lesker FTAKC060CM1). The coax wire is connected to the feedthrough one side and the antenna on the other side with inline barrel connectors (Kurt J. Lesker FTAIBC058). The antenna was built out of high uhv copper wire that was cut at the appropriate length. This antenna radiates the microwave equally in all azimuthal directions about the antenna. Therefore, we oriented our antenna parallel to the atoms to maximize the microwave radiation that the atoms can receive, see Fig. \ref{fig_5}.

\section{Performance}

\begin{figure}[t!]
\includegraphics[width=220pt]{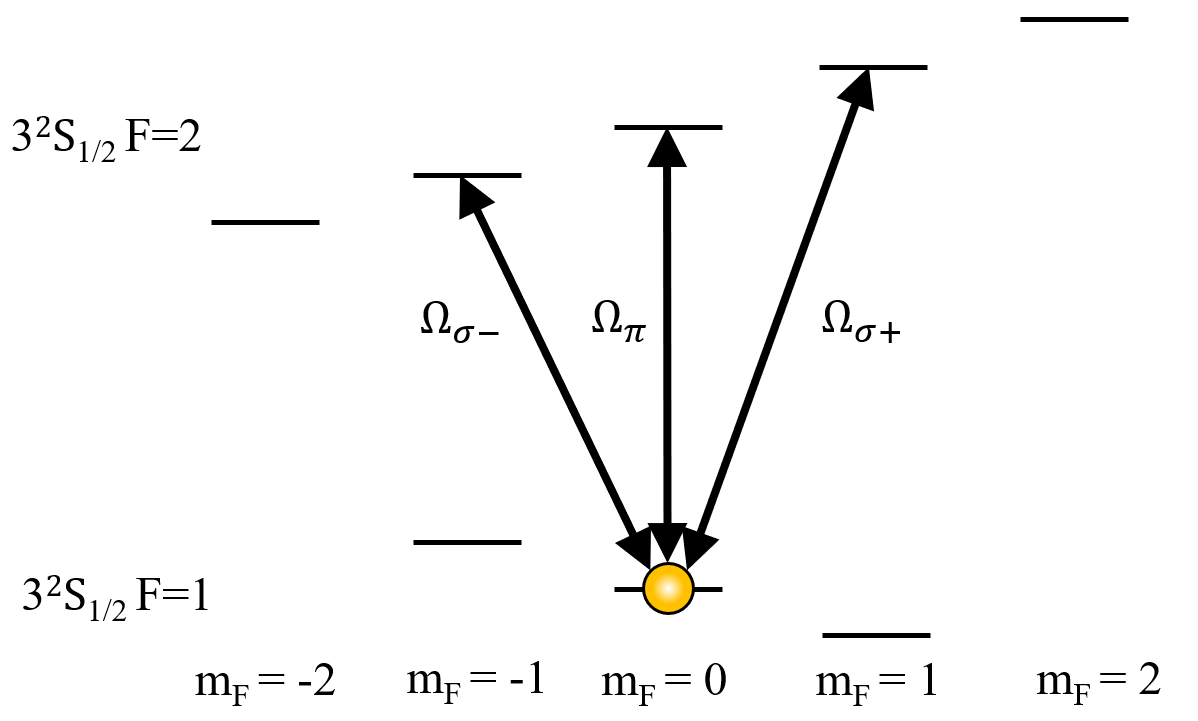}
\caption{(Color online) Na ground state hyperfine energy levels and associated Rabi couplings.}
\label{fig_6}
\end{figure}

\begin{figure}[t!]
\includegraphics[width=180pt]{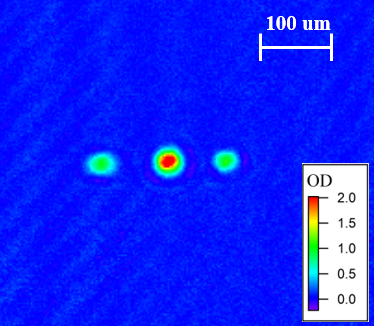}
\caption{(Color online) False-color Stern-Gerlach absorption image of spin populations in our Na spinor BEC, imaged after 7~ms time-of-flight expansion. The total atom number is N $\approx20,000$. The color scale represents optical density.}
\label{fig_7}
\end{figure}

To test the functionality of the microwave source, we excite Rabi oscillations in our Na spinor BEC in an applied B-field that gives a Zeeman shift of $\mu_B B/h = 300 \; \textrm{kHz}$, where $\mu_B$ is the Bohr magneton. Fig.~\ref{fig_6} shows the energy level diagram for the relevant transitions and Rabi oscillations. Here, $\Omega_\pi$ couples the $F=1$, $m_F=0$ level to the $F=2$, $m_F=0$ level, and similarly for $\Omega_{\sigma_-}$ and $\Omega_{\sigma_+}$. To find the resonance frequencies of the $\sigma_+$, $\sigma_-$ and $\pi$ transitions, shown as black arrows in Fig.~\ref{fig_6}, we measure the F = 1, $m_F=0$ population after a short, low-power microwave pulse, as a function of microwave frequency. The spin populations are measured using Stern-Gerlach time-of-flight absorption imaging, see Fig.~\ref{fig_7}.  

\begin{figure}[ht!]
\includegraphics[width=3.25in]{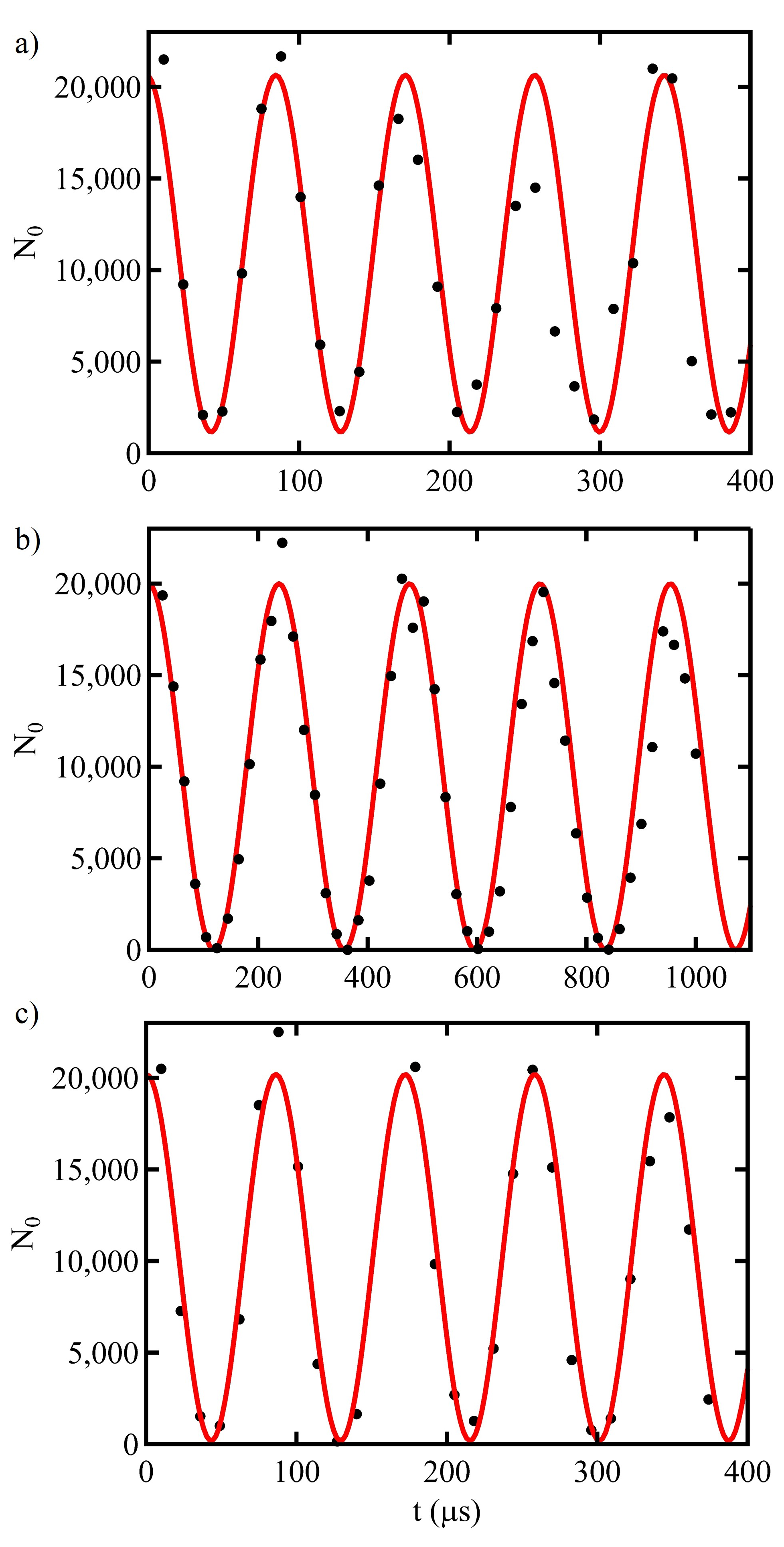}
\caption{(Color online) Rabi oscillations on resonance with the a) $\sigma_-$, b) $\sigma_\pi$ and c) $\sigma_+$ transitions. Shown is the measured number of atoms in F=1, $m_F=0$ (black circles) and a sinusoidal fit (red line) vs.~microwave pulse duration. The fit is used to extract the Rabi frequency a) $\Omega_{\sigma_-}$, b) $\Omega_{\sigma_\pi}$ and c) $\Omega_{\sigma_+}$. These measurements were done at the maximum microwave power of 20 W.}
\label{fig_8}
\end{figure}

Once the resonance frequencies are determined, we then produce microwaves at the resonant frequency with constant amplitude and maximum power (20 W) to maximize the Rabi frequencies. To deduce the Rabi frequencies, we measure the population remaining in F=1, $m_F=0$ as a function of pulse duration for on-resonant pulses, see Fig.~\ref{fig_8}.

We measured the Rabi frequencies to be
\begin{eqnarray*}
&\Omega_{\sigma-} = 2\pi \times (11.7 \pm 0.3 \; \textrm{kHz}) \\
&\Omega_\pi = 2\pi \times (4.1 \pm 0.2 \; \textrm{kHz}) \\
&\Omega_{\sigma+} = 2\pi \times (11.6 \pm 0.4 \; \textrm{kHz}) \\ 
\end{eqnarray*}
In Fig.~\ref{fig_8}, the black dots are the measured data and the red lines are sinusoidal fits. From these data, we conclude that our microwave source is working satisfactorily. We plan to use this control to implement devices in spin-based matter wave quantum optics, such as a quantum-enhanced interferometer and a phase-sensitive amplifier for matter waves.

\section{Supplemental Material}
The supplemental material folder contains the Verilog source code to run on a Terasic DE0-CV FPGA board, as well as sample LabVIEW code for communicating with the FPGA board to generate DDS output. Instructions on how to use the supplemental materials can be found in the README.txt file in the archive. Our Verilog source code was compiled in Quartus II version 14.0 on a computer running Windows 10.

\section{Conclusion}
We presented a versatile FPGA-based microwave source that is applicable in atomic physics experiments where control over the frequency and amplitude of the waves is desired. Our source was designed for experiments with microwave-dressed ultracold sodium gases, but it can be used to couple to other atomic transitions in the GHz regime, too. We tested our microwave source by producing Rabi oscillations in our Na spinor BEC with the microwaves on resonance with the transition frequencies of the ground state hyperfine manifold. Our design allows for versatile functionality of the FPGA. For example, with some changes to the existing code, the device could function as a servo controller or lock-in amplifier.

\vspace{-0.2in}
\section*{Acknowledgements}

We would like to thank Dr. Steven Olmschenk from Denison University, Ohio, for his input on this paper and for his original code that he shared with us. This material is based upon work supported by the National Science Foundation under Grant No.~1846965.

\end{document}